\documentclass{article}
\usepackage{graphicx}
\usepackage[super,comma,sort&compress]{natbib}

\newcommand*{\myaffil}[1]{\setcounter{footnote}{#1}\footnotemark\hspace{2mm}}

\begin{document}
\title{Influence of disorder on conductance in bilayer graphene under perpendicular electric field}
\author{Hisao Miyazaki\myaffil{1}\myaffil{2}, Kazuhito Tsukagoshi\myaffil{0}\myaffil{1}\myaffil{2}\myaffil{3}, \\ Akinobu Kanda\myaffil{2}\myaffil{4}, Minoru Otani\myaffil{3}, and Susumu Okada\myaffil{2}\myaffil{4}\myaffil{5}}
\date{}
\maketitle

\begin{center}
MANA,NIMS, Namiki, Tsukuba 305-0047, Japan, CREST, JST, Kawaguchi 332-0012, Japan, AIST, Higashi, Tsukuba 305-8562, Japan, Institute of Physics, University of Tsukuba, and Center for Computational Sciences, University of Tsukuba, Tsukuba 305-8571, Japan\\
\end{center}

\begin{center}
{\sf E-mail: TSUKAGOSHI.Kazuhito@nims.go.jp}
\end{center}

\setcounter{footnote}{1} \footnotetext{To whom corresponding should be addressed}
\setcounter{footnote}{2} \footnotetext{MANA, NIMS} 
\setcounter{footnote}{3} \footnotetext{CREST, JST}
\setcounter{footnote}{4} \footnotetext{AIST}
\setcounter{footnote}{5} \footnotetext{Institute of Physics, University of Tsukuba}
\setcounter{footnote}{6} \footnotetext{Center for Computational Sciences, University of Tsukuba}

\begin{abstract}
Electron transport in bilayer graphene placed under a perpendicular electric field is revealed experimentally. 
Steep increase of the resistance is observed under high electric field; however, the resistance does not diverge even at low temperatures. 
The observed temperature dependence of the conductance consists of two contributions: the thermally activated (TA) conduction and the variable range hopping (VRH) conduction. 
We find that for the measured electric field range (0 - 1.3 V/nm) the mobility gap extracted from the TA behavior agrees well with the theoretical prediction for the band gap opening in bilayer graphene, although the VRH conduction deteriorates the insulating state more seriously in bilayer graphene with smaller mobility. 
These results show that the improvement of the mobility is crucial for the successful operation of the bilayer graphene field effect transistor. 
\end{abstract}



 In the modern electronics, band gap is a key concept for switching current, and thus processing electric signals.\cite{sze} 
For graphene, an atomic layer of graphite, although it has great advantages for nanoelectronics applications, including atomically thin self-organized channel, high mobility,\cite{bolotin, morozov, du} good electrical contact with metals, feasibility of nanoscale fabrication,\cite{han} and resource abundance, its semimetallic electronic band structure makes it quite challenging for realization of the graphene transistor. 

In principle, band gap can be introduced to graphene by two methods: in the first, a graphene nanoribbon has a band gap due to the quantum confinement.\cite{son} 
Theoretically, the magnitude of the band gap depends on the channel width $W$ approximately as $\sim$ 1 eVnm$/W$ in armchair nanoribbons, while the band gap due to the confinement effect is absent in zigzag nanoribbons. 
In experiments, conduction suppression has been observed both in lithographically defined\cite{han} and chemically derived\cite{li} nanoribbons within an energy interval (transport gap) in agreement with the theoretical band gap, however, the dependence on the orientation (armchair or zigzag) was not observed. 
Consequently, the origin of the observed transport gap is controversial in association with quantum dot formation due to the surface disorder,\cite{sols, todd, liu} which is inevitably produced in the present nanofabrication processes. 

In the second, a band gap is predicted to generate in {\it bilayer} graphene  under a perpendicular electric field without any structural confinement.\cite{mccann, min, guo, otani} 
This is due to the breaking of the sublattice symmetry for the carbon atoms sitting at the A- and B- sites in the atomic structure. 
Such electric field is practically produced by chemical doping on graphene\cite{ohta, castro} and/or by a gate electric field.\cite{zhang, mak, oostinga, xia} 
So far, the band gap formation has been confirmed in optical spectroscopic measurements such as angle-resolved photoemission spectroscopy (ARPES)\cite{ohta} and infrared spectroscopy.\cite{zhang, mak, kuzmenko}
The observed band gap depended on the gate electric field and reached $\sim 200$ meV, which was consistent with the theoretical prediction. \cite{mccann, min} 
On the other hand, in transport measurements,\cite{oostinga} although an insulating behavior was observed with the increase of resistance in lowering temperature, the observed temperature dependence of resistance at low temperatures (below 50 K) was not the thermal activation type, $R \propto \exp(E_{\rm a}/2k_{\rm B}T )$, which is normally regarded as a direct evidence of the band gap opening with the magnitude of $E_{\rm a}$, but the variable-range-hopping type, $R \propto \exp ((T_{\rm h}/T)^{1/3})$, typically observed in strongly disordered two-dimensional systems ($k_{\rm B}$ is Boltzmann constant, $T$ is temperature, and $T_{\rm h}$ is a constant). 
The authors of Reference \raisebox{-1.7mm}{\Large \cite{oostinga}} claimed that the experimentally obtained band gap was below 10 meV, in spite of the fact that the device structure was similar to those used in Reference \raisebox{-1.7mm}{\Large \cite{zhang}}, which reports a band gap over 200 meV in the optical spectroscopic measurement. 
The origin of this missing band gap in transport measurement is not clarified so far, and this is one of the biggest issues in graphene research. 
In this paper, based on measurements of the gate-voltage and temperature dependence of transport properties, we make clear the origin of the missing band gap: we show that although the sign of a large band gap is present even in transport measurement, with a magnitude comparable to the theoretical prediction, the insulation within the band gap is more seriously deteriorated in bilayer graphene with lower mobility. 

For making a double-gated graphene transistor (Figure 1), we employed a simple method for formation of a top gate with a self-organized insulating layer.
The following is the fabrication procedure:
First, graphene films were placed on a highly-doped Si substrate with a 90-nm or 300-nm SiO$_2$ layer on top by micromechanical cleavage of Madagascar natural graphite.\cite{novoselov} 
The number of layers of a film was determined from the intensity of the green signal in an optical microscope image.\cite{oostinga} 
Next, two steps of electrode formation were carried out by using the $e$-beam lithography followed by the metal deposition: the first for the source and drain electrodes (Au (50 nm)/Ti (5 nm)) and the second for the top gate (30-nm Al). 
Then the device was exposed to the air for more than an hour. 
It is rather surprising that although the Al film was directly deposited on the graphene film, the exposure to the air eventually blocked the conduction between the Al layer and graphene. We attribute it to the formation of a self-organized Al-oxide layer between them.\cite{miyazaki, miyazakiSST}
The resulting leakage current from the Al film to the graphene channel was less than 1 nA, even after the device was stored in vacuum, so that we were able to use the Al film as a top gate. 
Thus, in our device configuration, the Al film behaves as a top-gate, while the highly doped substrate acts as a back-gate. 
We note that in the local top gate structure \cite{huard} shown in Figure 1, the top-gate voltage (\mbox{$V_{\rm tg}$}) affects the channel locally underneath the top gate, while the back-gate voltage (\mbox{$V_{\rm bg}$}) changes the carrier density over the whole area of the graphene channel. 

The devices were cooled with liquid helium or nitrogen, and the dc electron transport properties of the devices were examined for temperatures between 4K and 200K. 

Figure 2 shows the $V_{\rm tg}$ and $V_{\rm bg}$ dependence of the zero-bias resistance in monolayer, bilayer, trilayer and tetralayer devices. 
For all devices, there appear two ridge lines (dashed lines and dotted lines in Figure 2): one is diagonal in the $V_{\rm bg}-V_{\rm tg}$ plane and the other along the $V_{\rm tg}$-axis. 
The former corresponds to the charge neutrality of the graphene area which is sandwiched between the top and back gates (top-gate area), while the latter to the charge neutrality of the areas only affected by the back gate (i.e., not covered with the top gate: uncovered area).
We note that these two kinds of areas are connected in series in our device
configuration (see Fig. 1).
Thus, the intersecting point of the ridge lines represents the charge
neutrality of the whole graphene. 
A similar behavior has been commonly observed in double-gated monolayer graphene devices,\cite{huard} indicating that our double-gate structure works properly. 
The slope of the former ridge line $\alpha ={\rm d}V_{\rm tg}/{\rm d}V_{\rm bg}$ corresponds to the ratio of the back gate capacitance to the top gate capacitance per unit area, $\alpha =-(\epsilon_{\rm b} /d_{\rm b})/(\epsilon_{\rm t} /d_t)$, where $\epsilon_{\rm b(t)}$ and $d_{\rm b(t)}$ are the dielectric constant and the thickness of the back (top) gate, respectively.
The slope is $\alpha \approx -0.013$ for the substrate with a 300-nm SiO$_2$ layer ($d_{\rm b} = 300$ nm). 
By using the dielectric constants of the SiO$_2$ ($3.9\epsilon_0$, where $\epsilon_0$ is the dielectric constant of vacuum) and Al-oxide ($6\epsilon_0 -10\epsilon_0$) for the back and top gates, one can estimate the thickness of the top gate insulator to be $d_{\rm t} = 5-9$  nm, which is thinner than the reported values for the thickness of top gate dielectrics (Al$_2$O$_3$\cite{zhang}, SiO$_2$,\cite{oostinga} and so on), ensuring lower operating voltages for our devices.
It is noted that only for the bilayer device (Figure 2b), the resistance of the diagonal ridge line steeply increases with the gate electric field (i.e., the difference between the top and back gate voltages). 
This behavior was reproducibly observed in 39 out of 40 bilayer devices.\cite{note3} 
The peak resistance becomes higher at lower temperatures (shown below), reflecting a decrease of carrier density at the charge neutrality point, implying the band gap formation.
These results are consistent with the observation reported in Reference \raisebox{-1.7mm}{\Large \cite{oostinga}}.

The temperature dependence of the minimum conductance (i.e., on the diagonal ridge) at a fixed perpendicular electric field $E$ provides basic insight into the band gap opening in bilayer graphene. 
Figure 3a shows the typical temperature dependence of the conductance at the charge neutrality point (on the diagonal ridge) with an electric field $E = $ 1.2 V/nm. 
Here, from the elementary electrostatics, the electric field is given by  $E=\{ (V_{\rm tg}-V_{\rm tg}^0)-(V_{\rm bg}-V_{\rm bg}^0) \}/\{(\epsilon_0 / \epsilon_{\rm b}) d_{\rm b} + d_{\rm gr}+(\epsilon_0 / \epsilon_{\rm t}) d_t \}$, where $V_{\rm tg}^0$ and $V_{\rm bg}^0$ are the gate voltages which give the minimum resistance on the diagonal ridge line, corresponding to $E=0$, and $d_{\rm gr}$ is the thickness of the bilayer graphene, which is negligible in the expression for our device structure. 
As shown in the inset, the low temperature part is well explained by the variable range hopping (VRH) conduction for two-dimensional systems, $G_h \propto \exp \{- (T_{\rm h} /T)^{1/3} \}$ ($T_{\rm h}$ is a constant), which agrees with the results reported in Reference \raisebox{-1.7mm}{\Large \cite{oostinga}}, while at high temperatures deviation from the VRH conduction is remarkable. 
The overall characteristics can be well fitted by the sum of the VRH conduction and the thermally activated (TA) conduction, $G_{\rm a}\propto \exp (-E_{\rm a} /2k_{\rm B} T)$, as shown in the main panel of Figure 3a. 
The resultant values of the fitting parameters are $E_{\rm a} =$ 136 meV and $k_{\rm B} T_{\rm h} =$ 1.2 meV. 

It is known that this behavior (VRH + TA) is commonly observed in strongly disordered systems such as amorphous semiconductors. \cite{ambegaokar, nang} The activation energy is maximized when the Fermi level is placed at the middle of the mobility gap between the top of the extended (delocalized) state in the valence band (called mobility edge) and the bottom of the extended states in the valence band (mobility edge), and the maximum value corresponds to the magnitude of the mobility gap.
In disordered systems, the mobility gap is filled with the localized states and electron hopping across localized states contributes dominantly to the conduction inside the mobility gap at low temperatures. 
Here, a larger value of $T_{\rm h}$ corresponds to smaller contribution of the VRH to the conductance.
Such localized states are likely to be introduced to graphene by vacancies and impurities (see Figure 3b).\cite{pereira, nilsson}

The electric field dependence of $E_{\rm a}$ and $T_{\rm h}$ are plotted in Figure 3c. Both $E_{\rm a}$ and $T_{\rm h}$ increase with $E$, and surprisingly, the mobility gap $E_{\rm a}$ agrees well with the theoretical band gap $\delta_{\rm gap}^{\rm LDA}$ (dashed line in Figure 3c), which is calculated within the framework of the local density approximation (LDA).\cite{hohenberg, otani2} 
These results clearly shows that the band gap with reasonable values develops with electric field even in bilayer graphene samples for transport measurements, but the hopping conduction due to the disorder strongly deteriorates the insulation within the band gap.
Note that the slight increase of $T_{\rm h}$ with the electric field (Figure 3c, right panel) suggests that the localized states are also affected by the electric field. 
In the theory of the hopping conduction, $T_{\rm h}$ is inversely proportional to the density of states.\cite{ambegaokar} 
Thus, $T_{\rm h}$ in Figure 3c suggests that the number of the localized impurity states effectively decreases with increasing electric field, which agrees qualitatively with the theoretical prediction.\cite{nilsson}

 Also note that due to the peculiar structure of the energy spectrum of the biased bilayer graphene, the density of states (DOS) of ideal (clean) bilayer graphene under a strong perpendicular electric field has singularities at the band edge.
In the presence of disorder, these singularities smear and the band tails extend into the gap, leading to the renormalization of the band gap.\cite{nilsson, mkhiraryan}
Our observation that the magnitude of the mobility gap (in the presence of the disorder) agrees with that of the theoretical band gap for pristine graphene means that the states induced by the disorder is mostly localized (see Figure 3b), indicating that the disorder of the bilayer graphene samples is relatively small.

In order to study how the disorder influences the electron transport in bilayer graphene, we measured the temperature dependence of conductance for several samples with different sample qualities (reflected by mobility) at a fixed electric field $E = 1.3$ V/nm. 
Figure 4a shows an Arrhenius plot of the minimum conductance for samples with field-effect hole mobility of 460 cm$^2$/Vs and 1,100 cm$^2$/Vs. Clearly, the slope in low temperature part (VRH conduction) depends strongly on sample, whereas that of the higher temperature part (TA conduction) does not. 
The extacted $E_{\rm a}$ and $k_{\rm B} T_{\rm h}$ are summarized in Figure 4b as a function of the field-effect mobility $\mu$ of hole carrier for six samples. The mobility gap $E_{\rm a}$ ranges from 80 to 190 meV at $E=1.3$ V/nm and shows no considerable correlation with $\mu$. The values are close to the theoretical value $\delta_{\rm gap}^{\rm LDA}=$ 170 meV, indicating that the effect of disorder  on the mobility gap is relatively small in samples with mobility $\mu = $ 320 - 1100 cm$^2$/Vs. On the other hand, $T_{\rm h}$ increases with $\mu$, implying that bilayer graphene with higher mobility has lower density of the localized states. 

From this measurement, we can draw a conclusion that improvement of the graphene quality (mobility) is crucial to reduce the hopping conduction. 
A high electric field is also effective to suppress the hopping conduction, because $T_{\rm h}$ increases with the electric field (Figure 2b), as well as to enhance the band gap itself.
According to our LDA calculation, the band gap can be raised up to 280 meV, if $E = 5$ V/nm is applied to the bilayer graphene.\cite{otani} 
This value can be raised by decreasing the interlayer distance of bilayer.\cite{guo}

In conclusion, we investigate conduction of bilayer graphene under high electric fields, and show the existence of the mobility gap, which is comparable to the theoretical prediction of the band gap, in transport measurements. 
The temperature dependence of the minimum conductance is explained by the sum of band conduction and the variable range hopping conduction. 
A sequential analysis allow us to obtain the band gap of 80-190 meV at the electric field of 1.3 V/nm, which agrees with the LDA calculation $\sim 170$ meV. 
However, in transport measurements, the variable range hopping conduction across the localized states causes the current leakage, causing the degradation of the FET operation. 
We showed that raising the graphene mobility and/or applying higher electric fields is quite effective for reducing the influence of the variable range hopping conduction, and thus, improving the performance of the graphene FETs.

\section*{Acknowledgement.}
This study was supported in part by Grants-in-Aid for Scientific Research (Nos. 16GS50219, 17069004, and 18201028) from the Ministry of Education, Culture, Sports, Science and Technology of Japan. Some of the samples were prepared by S.-C. Choi. The authors would like to thank K. Novoselov and A. K. Geim for helpful discussions on the preparation of graphene flakes. Useful discussions with S.-L. Li, Y. Miyata, H. Kataura, K. Yanagi, and Y. Hayashi are also acknowledged. 

\bibliography{bilayergap}

\providecommand{\refin}[1]{\\ \textbf{Referenced in:} #1}
\begin{thebibliography}{10}

\bibitem{sze}
Sze,~S.~M.;\ \ K.,~N.~K. \textit{Physics of Semiconductor Devices;}
  Wiley-Interscience: New Jersey, 3rd ed.; 2006.

\bibitem{bolotin}
Bolotin,~K.~I.;\ \ Sikes,~K.~J.;\ \ Jiang,~Z.;\ \ Klima,~M.;\ \ Fudenberg,~G.;\
  \ Hone,~J.;\ \ Kim,~P.;\ \ Stormer,~H.~L. \textit{Solid State Commun..}
  \textbf{2008,} \textsl{146,} 351-355.

\bibitem{morozov}
Morozov,~S.~V.;\ \ Novoselov,~K.~S.;\ \ Katsnelson,~M.~I.;\ \ Schedin,~F.;\ \
  Elias,~D.~C.;\ \ Jaszczak,~J.~A.;\ \ Geim,~A.~K. \textit{Phys. Rev. Lett.}
  \textbf{2008,} \textsl{100,} 016602.

\bibitem{du}
Du,~X.;\ \ Skachko,~I.;\ \ Barker,~A.;\ \ Andrei,~E.~Y. \textit{Nat Nanotech.}
  \textbf{2008,} \textsl{3,} 491-495.

\bibitem{han}
Han,~M.~Y.;\ \ Oezyilmaz,~B.;\ \ Zhang,~Y.;\ \ Kim,~P. \textit{Phys. Rev.
  Lett.} \textbf{2007,} \textsl{98,} 206805.

\bibitem{son}
Son,~Y.-W.;\ \ Cohen,~M.~L.;\ \ Louie,~S.~G. \textit{Phys. Rev. Lett.}
  \textbf{2006,} \textsl{97,} 216803.

\bibitem{li}
Li,~X.;\ \ Wang,~X.;\ \ Zhang,~L.;\ \ Lee,~S.;\ \ Dai,~H. \textit{Science}
  \textbf{2008,} \textsl{319,} 1229-1232.

\bibitem{sols}
Sols,~F.;\ \ Guinea,~F.;\ \ Neto,~A. H.~C. \textit{Phys. Rev. Lett.}
  \textbf{2007,} \textsl{99,} 166803.

\bibitem{todd}
Todd,~K.;\ \ Chou,~H.-T.;\ \ Amasha,~S.;\ \ Goldhaber-Gordon,~D. \textit{Nano
  Lett.} \textbf{2009,} \textsl{9,} 416-421.

\bibitem{liu}
Liu,~X.;\ \ Oostinga,~J.~B.;\ \ Morpurgo,~A.~F.;\ \ Vandersypen,~L. M.~K.
  \textit{Phys. Rev. B} \textbf{2009,} \textsl{80,} 121407.

\bibitem{mccann}
McCann,~E. \textit{Phys. Rev. B} \textbf{2006,} \textsl{74,} 161403.

\bibitem{min}
Min,~H.;\ \ Sahu,~B.;\ \ Banerjee,~S.~K.;\ \ MacDonald,~A.~H. \textit{Phys.
  Rev. B} \textbf{2007,} \textsl{75,} 155115.

\bibitem{guo}
Guo,~Y.;\ \ Guo,~W.;\ \ Chen,~C. \textit{Appl.Phys. Lett.} \textbf{2008,}
  \textsl{92,} 243101.

\bibitem{otani}
Otani,~M.;\ \ Okada,~S. \textit{J. Phys. Soc. Jpn.} \textbf{2010,} \textsl{79,}
  073701.

\bibitem{ohta}
Ohta,~T.;\ \ Bostwick,~A.;\ \ Seyller,~T.;\ \ Horn,~K.;\ \ Rotenberg,~E.
  \textit{Science} \textbf{2006,} \textsl{313,} 951-954.

\bibitem{castro}
Castro,~E.~V.;\ \ Novoselov,~K.~S.;\ \ Morozov,~S.~V.;\ \ Peres,~N. M.~R.;\ \
  Dos~Santos,~J. M. B.~L.;\ \ Nilsson,~J.;\ \ Guinea,~F.;\ \ Geim,~A.~K.;\ \
  Neto,~A. H.~C. \textit{Phys. Rev. Lett.} \textbf{2007,} \textsl{99,} 216802.

\bibitem{zhang}
Zhang,~Y.;\ \ Tang,~T.-T.;\ \ Girit,~C.;\ \ Hao,~Z.;\ \ Martin,~M.~C.;\ \
  Zettl,~A.;\ \ Crommie,~M.~F.;\ \ Shen,~Y.~R.;\ \ Wang,~F. \textit{Nature}
  \textbf{2009,} \textsl{459,} 820-823.

\bibitem{mak}
Mak,~K.~F.;\ \ Lui,~C.~H.;\ \ Shan,~J.;\ \ Heinz,~T.~F. \textit{Phys. Rev.
  Lett.} \textbf{2009,} \textsl{102,} 256405.

\bibitem{oostinga}
Oostinga,~J.~B.;\ \ Heersche,~H.~B.;\ \ Liu,~X.;\ \ Morpurgo,~A.~F.;\ \
  Vandersypen,~L. M.~K. \textit{Nat. Mater.} \textbf{2008,} \textsl{7,}
  151-157.

\bibitem{xia}
Xia,~F.;\ \ Farmer,~D.~B.;\ \ Lin,~Y.-m.;\ \ Avouris,~P. \textit{Nano Lett.}
  \textbf{2010,} \textsl{10,} 715-718.

\bibitem{kuzmenko}
Kuzmenko,~A.~B.;\ \ van Heumen,~E.;\ \ van~der Marel,~D.;\ \ Lerch,~P.;\ \
  Blake,~P.;\ \ Novoselov,~K.~S.;\ \ Geim,~A.~K. \textit{Phys. Rev. B}
  \textbf{2009,} \textsl{79,} 115441.

\bibitem{novoselov}
Novoselov,~K.;\ \ Geim,~A.;\ \ Morozov,~S.;\ \ Jiang,~D.;\ \ Zhang,~Y.;\ \
  Dubonos,~S.;\ \ Grigorieva,~I.;\ \ Firsov,~A. \textit{Science} \textbf{2004,}
  \textsl{306,} 666-669.

\bibitem{miyazaki}
Miyazaki,~H.;\ \ Odaka,~S.;\ \ Sato,~T.;\ \ Tanaka,~S.;\ \ Goto,~H.;\ \
  Kanda,~A.;\ \ Tsukagoshi,~K.;\ \ Ootuka,~Y.;\ \ Aoyagi,~Y. \textit{Appl.
  Phys. Exp.} \textbf{2008,} \textsl{1,} 034007.

\bibitem{miyazakiSST}
Miyazaki,~H.;\ \ Li,~S.-L.;\ \ Kanda,~A.;\ \ Tsukagoshi,~K.
  \textit{Semiconductor Science and Technology} \textbf{2010,} \textsl{25,}
  034008.

\bibitem{huard}
Huard,~B.;\ \ Sulpizio,~J.~A.;\ \ Stander,~N.;\ \ Todd,~K.;\ \ Yang,~B.;\ \
  Goldhaber-Gordon,~D. \textit{Phys. Rev. Lett.} \textbf{2007,} \textsl{98,}
  236803.

\bibitem{note3}
We found that for bilayer graphene devides with a 300-nm SiO$_2$ layer, the
  resistance peak at the lower-right corner (with negative $V_{tg}$ and
  positive $V_{bg}$) becomes smaller than that at the upper-left corner, or
  sometimes even disappears. The origin of this asymmetry is not clear at this
  moment.

\bibitem{ambegaokar}
Ambegaokar,~V.;\ \ Halperin,~B.~I.;\ \ Langer,~J.~S. \textit{Phys. Rev. B}
  \textbf{1971,} \textsl{4,} 2612-2620.

\bibitem{nang}
Nang,~T.;\ \ Okuda,~M.;\ \ Matsushita,~T.;\ \ Yokota,~S.;\ \ Suzuki,~A.
  \textit{Jpn. J. Appl. Phys.} \textbf{1976,} \textsl{15,} 849-853.

\bibitem{pereira}
Pereira,~V.~M.;\ \ Guinea,~F.;\ \ Lopes~dos Santos,~J. M.~B.;\ \ Peres,~N.
  M.~R.;\ \ Castro~Neto,~A.~H. \textit{Phys. Rev. Lett.} \textbf{2006,}
  \textsl{96,} 036801.

\bibitem{nilsson}
Nilsson,~J.;\ \ Castro~Neto,~A.~H. \textit{Phys. Rev. Lett.} \textbf{2007,}
  \textsl{98,} 126801.

\bibitem{hohenberg}
Hohenberg,~P.;\ \ Kohn,~W. \textit{Phys. Rev.} \textbf{1964,} \textsl{136,}
  B864-871.

\bibitem{otani2}
Otani,~M.;\ \ Sugino,~O. \textit{Phys. Rev. B} \textbf{2006,} \textsl{73,}
  115407.

\bibitem{mkhiraryan}
Mkhitaryan,~V.~V.;\ \ Raikh,~M.~E. \textit{Phys. Rev. B} \textbf{2008,}
  \textsl{78,} 195409.

\end{thebibliography}
\bibliographystyle{achemso}

\begin{figure}[p]
\begin{center}
\includegraphics[width=100mm]{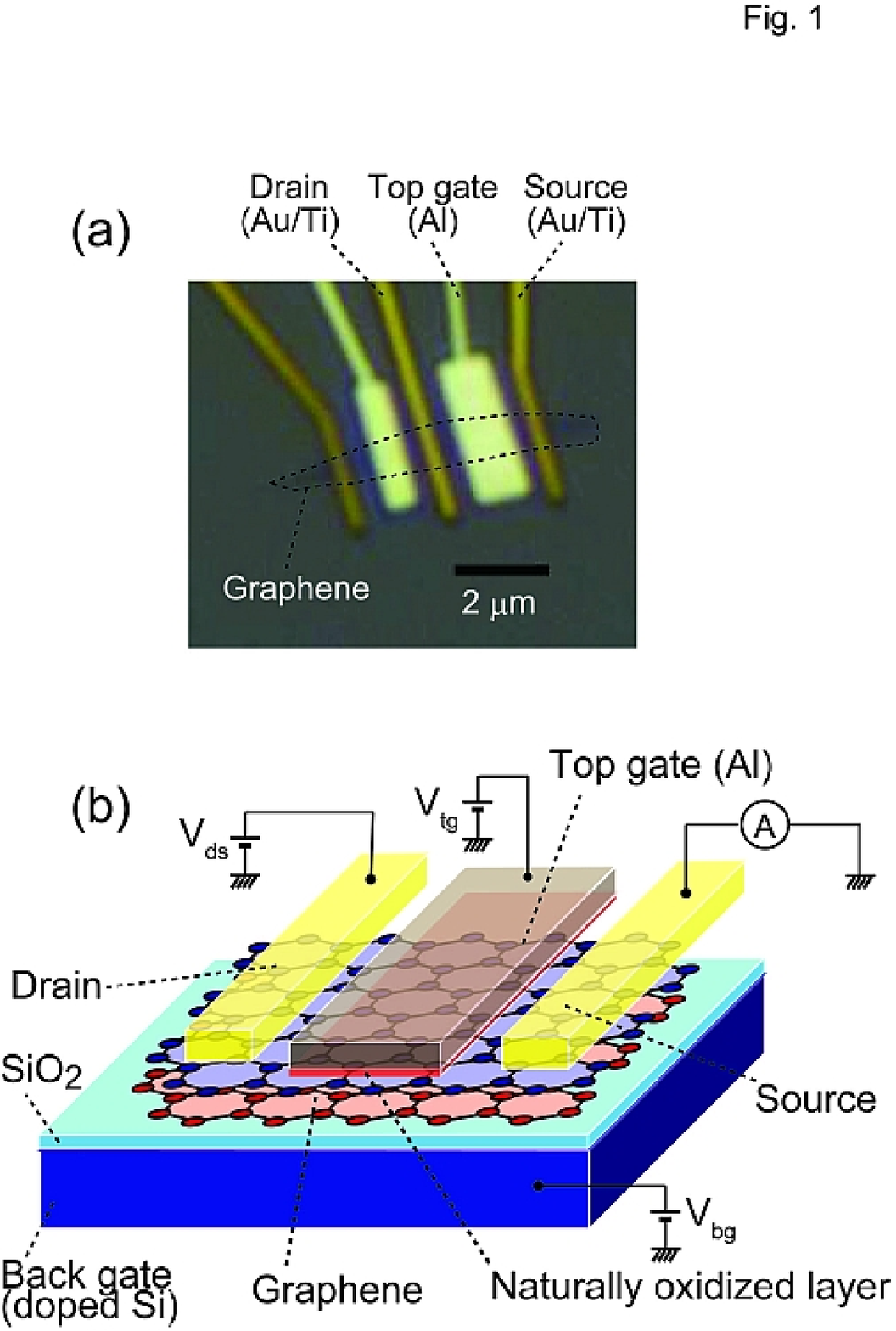}
 \end{center}
\caption{Optical microscope image (a) and schematic view (b) of a graphene transistor gated with SiO$_2$/Si-substrate back gate and Al top gate.} 
\end{figure}

\begin{figure}[p]
\begin{center}
  \includegraphics[width=100mm]{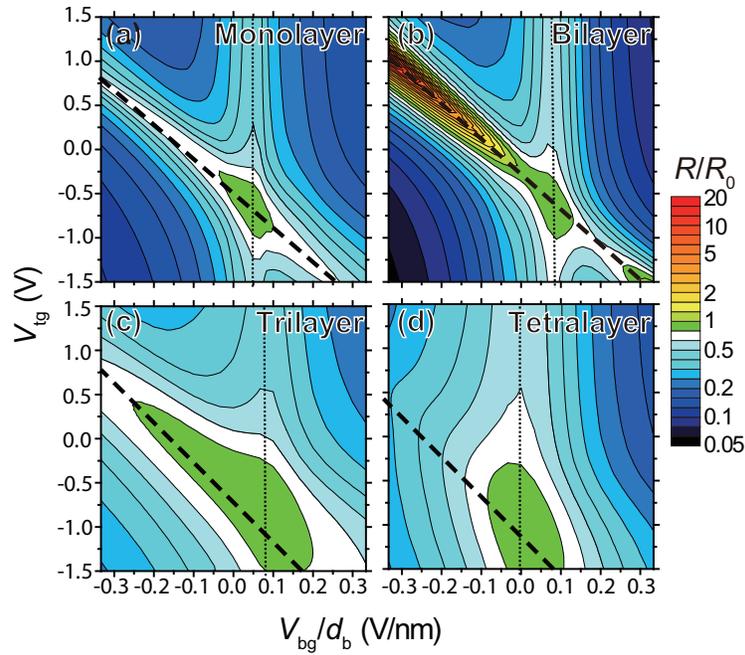}
 \end{center}
\caption{Contour plot of resistance as a function of the back-gate ($V_{\rm bg}$) and the top-gate ($V_{\rm tg}$) voltages for monolayer (a), bilayer (b), trilayer (c), and tetralayer (d) graphene. The bilayer sample (b) is on a Si substrate with a 90-nm SiO$_2$ layer, while the others (a, c, d) are on substrates with a 300-nm SiO2 layer.\cite{note3} The back-gate voltage is normalized by the thickness of the SiO$_2$ layer ($d_{\rm b}$). In each graphs, charge neutrality resistance peaks in two kinds of areas are indicated by dashed line (top-gated area) and dotted line (uncovered area).
The resistance is normalized by the resistance ($R_0$) at the point where the two peak lines cross.}
\end{figure}

\begin{figure}[p]
\begin{center}
  \includegraphics[width=100mm]{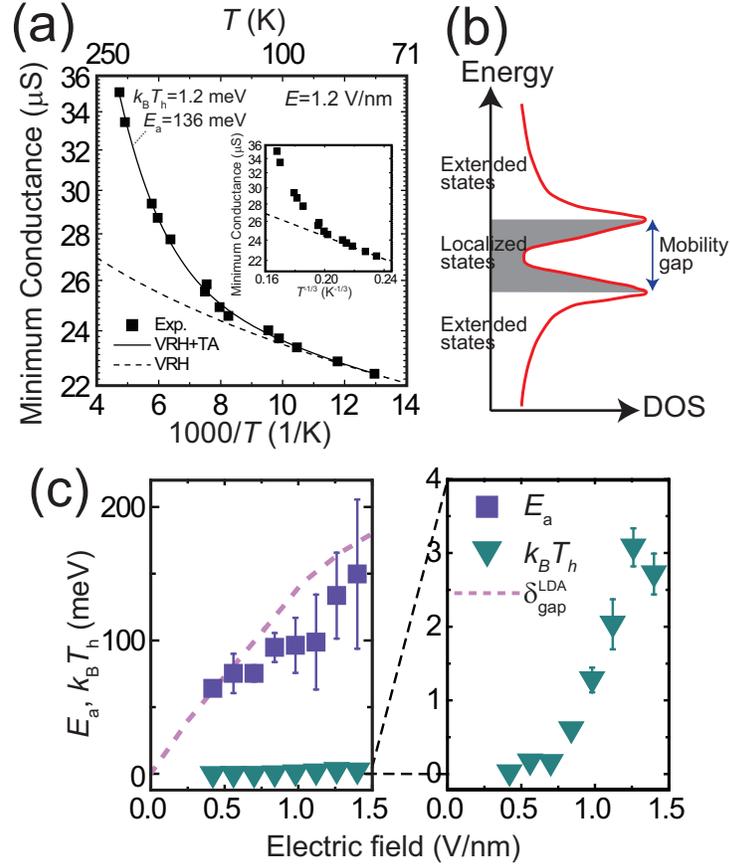}
 \end{center}
\caption{(a) Arrhenius plot of the minimum conductance at an electric field  $E=1.2$ V/nm. Experimental data (squares) are fitted by a sum of the variable-range-hopping (VRH) conduction and the thermally activated (TA) conduction (solid curve). The contribution by the VRH conduction is indicated by a dashed curve. Inset is a semilog plot of the same data as a function of $T^{-1/3}$. (b) Schematic view of the density of states (DOS) of disordered bilayer graphene under a perpendicular electric field.
(c) Thermal activation energy $E_{\rm a}$ for TA conduction (square) and characteristic energy $k_{\rm B}T_{\rm h}$ for VRH conduction (triangle) as a function of the electric field. A band gap calculated with the local density approximation (LDA) is shown by the dashed curve.}
\end{figure}

\begin{figure}[p]
\begin{center}
  \includegraphics[width=100mm]{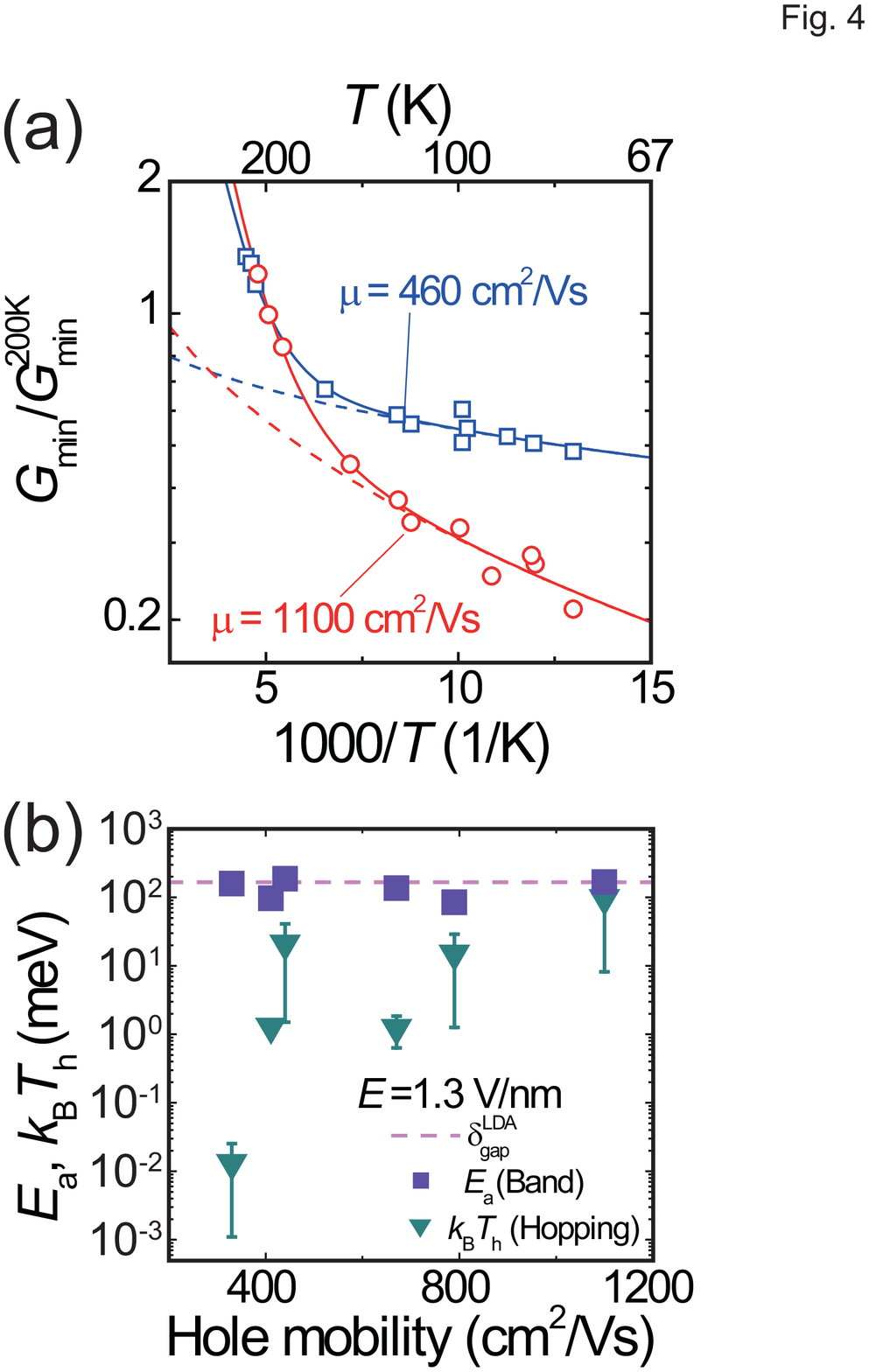}
 \end{center}
\caption{(a) Arrhenius plot of the minimum conductance for samples with the field-effect hole mobility ($\mu$) of 460 cm$^2$/Vs and 1,100 cm$^2$/Vs. The conductance is normalized by a value at $T=200$ K. Results of the fitting by the TA+VRH model (solid curve) and a contribution by VRH (dashed curve) are also shown. (b) Thermal activation energy $E_{\rm a}$ for TA conduction (square) and characteristic energy $k_{\rm B}T_{\rm h}$ for VRH conduction (triangle) for six samples as a function $\mu$ with the electric field of $E=1.3$ V/nm. A band gap calculated by the LDA is also shown (dashed line).}
\end{figure}

\end{document}